\documentclass{article}

\usepackage{arxiv}

\usepackage[utf8]{inputenc} 
\usepackage[T1]{fontenc}    
\usepackage{hyperref}       
\usepackage{url}            
\usepackage{booktabs}       
\usepackage{amsfonts}       
\usepackage{nicefrac}       
\usepackage{microtype}      
\usepackage{lipsum}		
\usepackage{graphicx}
\usepackage{natbib}
\usepackage{doi}
\usepackage{caption}
\usepackage{subcaption}
\usepackage{algorithm}
\usepackage{algorithmic}

\title{Circular Average Filtering and Circular Linear Interpolation in Complex Color Spaces}

\author{ \href{https://orcid.org/0000-0003-3618-4166}{\includegraphics[scale=0.06]{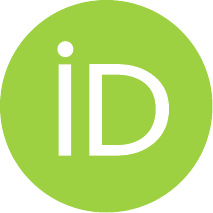}\hspace{1mm}Ergun Akleman}\thanks{Joint with Computer Science and Engineering. (webpage, alternative address)---\emph{not} for acknowledging funding agencies.} \\
	Visual Computing \& Computational Media,\\ Texas A\&M University, College Station, TX, 77831\\
	\texttt{ergun@tamu.edu} \\
	\And
	\href{https://orcid.org/0009-0000-0640-1882}{\includegraphics[scale=0.06]{orcid.pdf}\hspace{1mm}Shubham Agarwall} \\
	Computer Science and Engineering, \\ Texas A\&M University, College Station, TX, 77831\\
	\texttt{cherrycodes@tamu.edu} \\
     \And
	\href{https://orcid.org/0000-0001-9662-4640}{\includegraphics[scale=0.06]{orcid.pdf}\hspace{1mm}Donald H. House} \\
	Visual Computing \& Computational Media, \\ Texas A\&M University, College Station, TX, 77831\\
	\texttt{d-house@tamu.edu} \\
     \And
	\href{https://orcid.org/0000-0002-4834-9389}{\includegraphics[scale=0.06]{orcid.pdf}\hspace{1mm}Tolga Talha Yildiz} \\
	Computer Science and Engineering, \\ Texas A\&M University, College Station, TX, 77831\\
	\texttt{tolgayildiz@tamu.edu} \\
}

\renewcommand{\shorttitle}{Circular Average in Complex Color Spaces}

\hypersetup{
pdftitle={A template for the arxiv style},
pdfsubject={q-bio.NC, q-bio.QM},
pdfauthor={David S.~Hippocampus, Elias D.~Striatum},
pdfkeywords={First keyword, Second keyword, More},
}

\begin{document}
\maketitle

\begin{abstract}
In color spaces where the chromatic term is given in polar coordinates, the shortest distance between colors of the same value is circular. By converting such a space into a complex polar form with a real-valued value axis, a color algebra for combining colors is immediately available. 
In this work, we introduce two complex space operations utilizing this observation: circular average filtering and circular linear interpolation. These operations produce Archimedean Spirals, thus guaranteeing that they operate along the shortest paths.
We demonstrate that these operations provide an intuitive way to work in certain color spaces and that they are particularly useful for obtaining better filtering and interpolation results. We present a set of examples based on the perceptually uniform color space CIELAB or L*a*b* with its polar form CIEHLC. We conclude that representing colors in a complex space with circular operations can provide better visual results by exploitation of the strong algebraic properties of complex space $\mathbb{C}$.
\end{abstract}

\keywords{color spaces \and complex algebra \and color interpolation \and color filtering \and color algebras}

\section{Introduction}

\begin{figure}
    \begin{subfigure}[t]{0.24\textwidth}
        \includegraphics[width=1.0\textwidth]{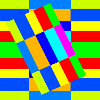}
        \includegraphics[width=1.0\textwidth]{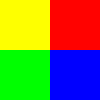}
        \caption{Original $100 \times 100$ Images.}
        \label{fig:filters2/coloredMsm23_originalDetail.png}
    \end{subfigure}
        \begin{subfigure}[t]{0.24\textwidth}
        \includegraphics[width=1.0\textwidth]{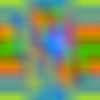} 
        \includegraphics[width=1.0\textwidth]{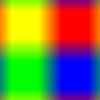}
        \caption{23x23 standard box filters in RBG.}
        \label{fig:filters2/coloredMsm23_rgbDetail.png}
    \end{subfigure}
      \hfill
    \begin{subfigure}[t]{0.24\textwidth}
        \includegraphics[width=1.0\textwidth]{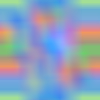} 
        \includegraphics[width=1.0\textwidth]{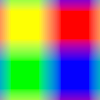}
        \caption{23x23 standard box filters in CIELAB.}
        \label{fig:filters2/coloredMsm23_labDetail.png}
    \end{subfigure}
      \hfill
          \begin{subfigure}[t]{0.24\textwidth}
        \includegraphics[width=1.0\textwidth]{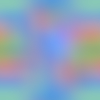}
         \includegraphics[width=1.0\textwidth]{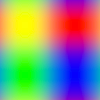}
        \caption{23x23 circular average box filters in CIEHLC.}
        \label{fig:filters2/coloredMsm23_hclDetail.png}
    \end{subfigure}
      \hfill
\caption{A comparison of standard convolution filters in RGB and CIELAB with circular average filters in CIEHLC. Note that circular average filters in CIEHLC consistently provide smoother results. }
\label{fig:filtercomparisons2Detailed}
\end{figure}

Large design spaces that provide expressive power and flexibility are usually created by using different types of algebras and operations to provide a wide range of design choices. For instance, in shape modeling matrix and vector algebra, quaternion algebra \cite{shoemake1985animating}, Clifford algebras \cite{hildenbrand2022geometric,vince2008geometric,hestenes2000geometric}, homogeneous coordinates \cite{roberts1963machine,bloomenthal1994homogeneous}, and barycentric algebras \cite{bartels1995introduction,farin1983algorithms,piegl1996nurbs,prautzsch2002bezier} are all used, each of which provides different types of design choices.

Similarly, in working with color we would like to have a variety of color manipulation design choices. However, with colors, there is no similar variety of algebraic approaches to choose from. For instance, colors in RGB are represented as a set of positive real numbers, which makes it hard to use anything other than barycentric algebras. The main reason behind this problem is that without negative numbers we do not have inverse elements. Important exceptions are the perceptual color spaces that are homologically similar to CIELAB \cite{joblove1978color,bora2015comparing}, such as CIELUV or OKLAB \cite{ottosson2020}, where the chromatic dimensions $a$ and $b$ can be any real number. 

We have recently realized that in these spaces chroma can be considered as a complex number in the form $a+ib$. Because of the strong algebraic properties of complex space $\mathbb{C}$ this interpretation immediately gives us an algebraic system. 
Another important implication of considering $a$ and $b$ as the real and imaginary components of a complex number is that CIEHLC (or alternatively $CIELCh_{uv}$) simply becomes the complex polar form of CIELAB. In this way, we do not have to view them as polar coordinates but we can operate in CIEHLC using the underlying complex space. It is straightforward to rotate and scale in this space using complex multiplication. Another big advantage is that the multiplicative inverse also exists. We will also demonstrate that working in complex space helps to formalize computing shortest angular paths. 

\begin{figure}
  \centering
    \begin{subfigure}[t]{0.32\textwidth}
        \includegraphics[width=1.0\textwidth]{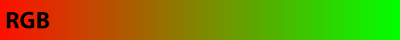}
        \includegraphics[width=1.0\textwidth]{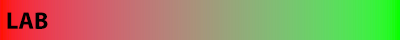}
        \includegraphics[width=1.0\textwidth]{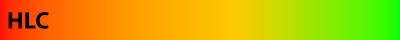}
        \caption{\it FF0F00 to 00FF00.}
        \label{fig:Linear/0}
    \end{subfigure}
      \hfill
    \begin{subfigure}[t]{0.32\textwidth}
        \includegraphics[width=1.0\textwidth]{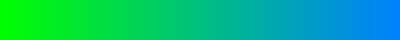}
        \includegraphics[width=1.0\textwidth]{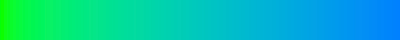}
         \includegraphics[width=1.0\textwidth]{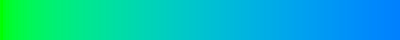}
        \caption{\it FF000F to 00FF00.}
        \label{fig:Linear/1}
    \end{subfigure}
      \hfill
    \begin{subfigure}[t]{0.32\textwidth}
        \includegraphics[width=1.0\textwidth]{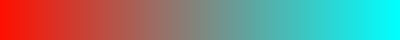}
        \includegraphics[width=1.0\textwidth]{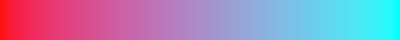}
        \includegraphics[width=1.0\textwidth]{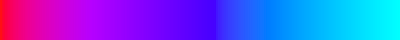}
        \caption{\it FF0F00 to 00FFFF. }
        \label{fig:Linear/2}
    \end{subfigure}
      \hfill
        \begin{subfigure}[t]{0.32\textwidth}
        \includegraphics[width=1.0\textwidth]{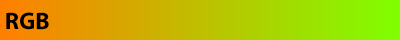}
        \includegraphics[width=1.0\textwidth]{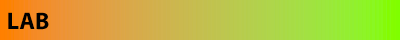}
        \includegraphics[width=1.0\textwidth]{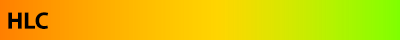}
        \caption{\it FF7F00 to 7FFF00.}
        \label{fig:Linear/4}
    \end{subfigure}
      \hfill
          \begin{subfigure}[t]{0.32\textwidth}
        \includegraphics[width=1.0\textwidth]{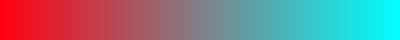}
        \includegraphics[width=1.0\textwidth]{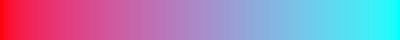}
        \includegraphics[width=1.0\textwidth]{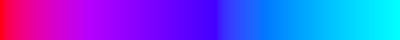}
        \caption{\it FF000F to 00FFFF.}
        \label{fig:Linear/3}
    \end{subfigure}
      \hfill
     \begin{subfigure}[t]{0.32\textwidth}
        \includegraphics[width=1.0\textwidth]{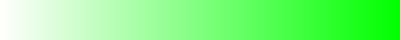}
        \includegraphics[width=1.0\textwidth]{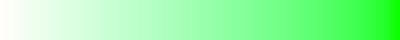}
        \includegraphics[width=1.0\textwidth]{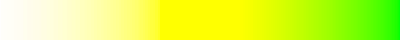}
        \caption{\it FFFEFD to 00FF00.}
        \label{fig:Linear/6}
    \end{subfigure}
      \hfill
\caption{Comparison of linear interpolations (LERP) in RGB and CIELAB with circular linear interpolation (CLERP) in CIEHLC. A key problem with LERP is that it tends to neutralize colors between the two control colors. Also, in RGB there can be a darkening of perceived value during the transition. CLERP in CIEHLC provides the shortest paths between hues. In this case, interpolations stay highly chromatic even when one control color is unsaturated as in (f), and perceived value transitions monotonically.
}
\label{fig:linearinterpolationcomparisons}
\end{figure}

Using standard RGB operations for baseline comparison, in this paper we focus on two types of color operations (interpolation and filtering) in two color spaces (RGB and CIELAB). We demonstrate that the complex algebras of these spaces are useful even when only using barycentric operations. It is recognized that interpolation and filtering operations in RGB can have a "darkening" effect on intermediate colors. On the other hand, in CIELAB it is possible to provide perceptually consistent interpolations and filters. In this paper, we introduce two circular operations for interpolation and filtering. We further demonstrate that the polar form of CIELAB, CIEHLC, can guarantee robust and consistent results with circular operations after the identification of the shortest circular paths in the corresponding CIELAB complex space. Our implementations further confirm that circular interpolation in CIEHLC provides even better results than linear interpolation in CIELAB since its strong perceptual properties are improved by following circular shortest paths. Figures~\ref{fig:filtercomparisons2Detailed} and \ref{fig:linearinterpolationcomparisons} preview the properties of these operations in comparison with standard filtering and interpolation approaches.

\subsection{Basis \& Rationale}

Barycentric operations in RGB space are one of the most common color operations in computer graphics applications. For example, they are used to obtain Gouraud-type \cite{gouraud1971continuous} interpolations to fill polygons from their corner colors \cite{floater2003mean,farbman2009coordinates}, and they are used in convolution filtering \cite{dudgeon1984multidimensional}. One major problem with barycentric operations in the RGB domain is that values (i.e. perceived luminance) of colors, as well as the chromatic saturation of colors, can decrease significantly when filtering between near complementary colors (See the RGB images in Figure~\ref{fig:filtercomparisons2Detailed}). 
One straightforward solution to preserving color value appears to be to implement barycentric operations directly in CIELAB  color space. In these spaces, it is possible to obtain better filtering as shown in Figure~\ref{fig:filtercomparisons2Detailed}. However, CIELAB also suffers from a conceptually similar problem. Standard convolution filters do not produce average values along the shortest paths in the color space. Our hypothesis was that to have better results we should compute values on the shortest paths defined by polar forms.

Despite their potential, it is not possible to have intuitive operations directly in polar forms such as CIEHLC \cite{zucconi2013}. As an example, consider two colors in polar form with angles $360^{\circ}-\delta$ and $\delta$, where $\delta$ is a small positive real number. Both of these colors are expected to be almost identical. However, their arithmetic average would be $180^{\circ}$, which is their complementary color. This problem, of course, can be solved by using the fact that $360^{\circ}-\delta=-\delta$. Then, the average gives us $0^{\circ}$, which is the expected result. However, this is clearly an ad hoc solution. 

In mathematics, the standard method to compute angular averages is called circular mean or circular average \cite{jammalamadaka2001topics,hotz2013extrinsic}. Each angular position is represented as a 2D unit vector. The direction of the averages of these unit vectors is considered the circular average. This method is useful for computing the circular average for two-dimensional circular data. The color systems that we normally work with, on the other hand, are at least three-dimensional, and not all dimensions are circular. Therefore, there is a need for circular operations that can work on color spaces. The main theoretical contribution of this paper is to provide a formal solution to this problem for color spaces by using complex spaces and identifying the shortest angular hue paths.

The problem of working with color spaces comes from the fact that changing hue is essentially a rotation operation and only rotation operators can guarantee to stay on the shortest path. The key property of the rotation operator is its non-commutativity. This is, in fact, also the main source of the problem in computing the average of two hues. Below, we show that this average corresponds to the geometric average in complex numbers. The geometric average on complex numbers is also not commutative and therefore it has two legal solutions based on the order of the operands, and one of them is not the shortest path. The solution to this problem is to restrict rotations to the actual shortest paths. We demonstrate that such a restriction can provide consistent interpolation and filtering operations in CIEHLC space.

\subsection{Overview}

There has been a significant amount of work on developing color spaces that provide perceptually consistent color representations \cite{seim1986towards,kuehni1999towards, ramanath2004spectral,luo2006uniform}, with the two primary considerations being that the axes provide perceptually independent directions in the space, and that perceptual color changes vary uniformly with position changes. Since the original Munsell color system \cite{nickerson1940history}, a common theme in color space construction has been to use some sort of polar form to represent colors \cite{zucconi2013,ottosson2020}. In this paper, however, we are not as interested in the ``geometric'' properties of any particular color space as we are in their homologic properties.  
Our basic insight is that most of the common color spaces, with the notable exception of RGB, can be represented by using a complex space along with a positive real axis. For instance, we view CIELAB as a complex space with chromatic axes $a$ and $b$, along with a third dimension of positive real numbers, which provides luminance $L$. 

The major advantage of using complex space is that a wide variety of well-behaved functions can be defined in this space, adding new options to our repertoire of color operations. In complex space, distances are well-defined, and we can also obtain the shortest angular paths using algebraic operations so that any operation defined on complex numbers will behave predictably. 
  
\subsection{Challenges \& Approach}

Although the potential for defining color operations is wide, in this paper we introduce our approach by focusing mainly on weighted averages of complex numbers. Computing the arithmetic average over complex numbers is straightforward and can be implemented in a straightforward way. There are two advantages: (1) they are commutative, and (2) any barycentric operation can be expressed as a successive application of arithmetic averages. Therefore, it is possible to include any barycentric formula such as blur filters or bilinear interpolation with minimal effort. As a consequence of these properties, the resulting barycentric operations are robust. They produce predictable color changes. They also guarantee that values of computed colors stay inside of the convex hull described by control colors in CIELAB. These operations visually act like an addition operation. Despite these advantages, using arithmetic average-based approaches have the problem that they do not necessarily produce points on the shortest paths. 

We have also developed geometric average-based approaches in CIELAB to provide a formal solution to this shortest-path problem. These averages can help to operate on polar forms such as CIEHLC color space. An advantage of geometric average-based approaches is that they can also preserve saturation. However, geometric average-based approaches are not that straightforward to implement. The difficulty comes from the fact that these are essentially SLERP operations \cite{shoemake1985animating,kim1995general} and are not commutative. There is, therefore, a need for solutions that do not produce sudden jumps in color. To develop a formal solution we have used a common property of complex numbers to identify the shortest path between two complex numbers. Using this formalism, we have developed a methodology to operate on polar forms to obtain predictable results. In the next section, we present a few extensions to complex number representations to make geometric average-based approaches feasible. We also show why there is not always a unique solution with geometric average-based approaches.

\section{Using Complex Space for Colors}
\label{sec:cvcolorspace}

We observe that a complex representation is natural for CIELAB since the two elements of CIELAB, $a$ and $b$, can be directly turned into a complex number as $a+bi$. Since $a$ and $b$ can be any real number, this transformation provides all the algebraic properties of complex space. We, furthermore, assume that this space is not bounded by $z\bar{z} \geq 1$ and it includes all $\mathbb{C}$ as in CIELAB space. The numbers outside of the range may not necessarily correspond to displayable colors, but this is not really essential. We can still make computations. If the final result is not in the displayable gamut we can always display it by finding the closest displayable color. But, while using our operations, we never have to make such a conversion, it is a display consideration not an algebraic consideration. This is the power of the algebrization process, it allows us to manipulate colors with any analytical functions. 

Despite the general power of the algebrization process, in this paper, we only focus on relatively simple operations. Let us first check two types of averages on complex numbers: arithmetic and geometric. These
two types of averages define two types of interpolations on complex numbers as follows
\begin{equation}
z(t) = z_0(1-t) + z_1 t  \label{Equ:ai}
\end{equation}
\begin{center} and \end{center}
\begin{equation}
z(t) = z_0^{(1-t)} z_1^{t}\label{Equ:gi}
\end{equation}
Equation~\ref{Equ:ai} is arithmetic or barycentric interpolation with parameter $0 \le t \le 1$. It produces a line segment in complex space. Equation~\ref{Equ:gi} is geometric interpolation or SLERP again with parameter $0 \le t \le 1$. It produces a segment of a logarithmic spiral in complex space. These two types of operations provide the basic framework for the two types of color operations discussed in this paper. 

The arithmetic average operation is based on operations that are straightforward to implement in $\mathbb{C}$, since distances are well defined. All barycentric operations that work on vectors can also work on complex numbers with no changes.  

\subsection{Geometric Averages in Complex Space}
To include SLERP type operations on complex numbers there is a need for using some special properties of complex numbers. There is also a need for making a set of minor updates. Most of the discussion and extensions in this section are to implement SLERP type operations on complex numbers. These extensions are useful since combining SLERP operations with CIEHLC interpolations provides an alternative approach to interpolating colors in polar forms. In this paper, we give examples in CIEHLC, but the same mathematical framework will also work with any polar form such as HSV or CIEHCL.

The geometric average operation can be obtained from the SLERP formula (Equation~\ref{Equ:gi}) by choosing $t=1/2$. Unlike the geometric average of real numbers, the geometric average of complex numbers given in Equation~\ref{Equ:gi} is not unique. This can be demonstrated with a simple example. Consider the following geometric average of two complex numbers:
$z_0^{1/2}  z_1^{1/2} = \sqrt{z_0 z_1}$. Let $z_0 = r_0 e^{i \theta_0}$ $z_1 = r_1 e^{i \theta_1}$. There are two square roots of this product given by
$$z_0^{1/2} = \{ r_0^{1/2} e^{i \frac{\theta_0}{2}}, r_0^{1/2} e^{i (\frac{\theta_0}{2} + \pi)} \}$$
\begin{center} and \end{center}
$$z_1^{1/2} = \{ r_1^{1/2} e^{i \frac{\theta_1}{2}}, r_1^{1/2} e^{i (\frac{\theta_1}{2} + \pi)} \}.$$
All possible multiplications of these two numbers give us again two complex numbers
\begin{equation}
z_0^{1/2} z_1^{1/2} = \{ \sqrt{r_0 r_1} \; \; e^{i \frac{\theta_0+\theta_1}{2}}, \sqrt{r_0 r_1} \; \; e^{i \frac{\theta_0+\theta_1+ 2\pi}{2}  } \} 
\end{equation}
This is exactly the problem of polar forms, which we casually discussed previously. 
Therefore, the geometric mean of two nearly identical colors can legally be two different colors. If the operation is not commutative, either one of the legal solutions can be chosen based on the input. In order to differentiate between the two, we need a formalism that is commonly used in vector, quaternion, and geometric algebras. For instance, assuming that we have a consistent right-hand or left-hand rule, the cross product of two vectors gives us a perpendicular vector along with a well-defined shortest angular path between the two vectors. We always rotate from the first vector (or quaternion) to the second in a given counterclockwise (or clockwise) direction. The same formalism also exists in complex algebra that we can exploit for color applications. 

\subsection{Finding Shortest Angular Paths}

One of the advantages of complex numbers is that it is straightforward to find the shortest angular path in $\mathbb{C}$. Consider the following multiplication of one complex number $z_0$ with the conjugate of another $z_1$:  
\begin{eqnarray}
z(t) = z_0 \bar{z}_1 &= & r_0 e^{i \theta_0} r_1 e^{-i \theta_1} \nonumber \\
&= & (r_0 cos \theta + i r_0 sin \theta) (r_1 cos \theta - i r_1 sin \theta \nonumber \\
&= & (x_0 + i y_0) (x_1 - i y_1) \nonumber \\
&= & x_0 x_1 + y_0 y_1 + i ( x_0 y_1 - y_0 x_1) \nonumber \\
&= &  r_0 r_1 e^{i \delta \theta} \nonumber \label{Eq:vector}
\end{eqnarray}
This multiplication corresponds to geometric algebra multiplication. The first term is the inner product or scalar multiplication, and the second term is the outer product or vector multiplication. This multiplication always chooses the shortest path between the two complex numbers regardless of how the original angles are selected. In this case, $\delta \theta$ is computed as a number between $-\pi$ and $\pi$. Positive $\delta \theta$ corresponds to counterclockwise rotation and negative $\delta \theta$ corresponds to clockwise rotation. Note that $\delta \theta$ can be uniquely computed. Using this shortest path, we can now choose 
$$\theta_1 = \theta_0 + \delta \theta$$
This guarantees that we have a unique solution that corresponds to the shortest path between two angles except for exactly complementary colors such as red and cyan. In such cases, $\delta \theta$ can be either $\pi$ or $-\pi$ and there will still be an ambiguity. This can be resolved by deciding on a rubric to choose either $-\pi \leq \delta \theta < \pi$ or $-\pi < \delta \theta \leq \pi$. 

By first establishing the shortest path we can define a unique geometric average. Moreover, this geometric average now is commutative. In other words
$\sqrt{z_0 z_1} = \sqrt{z_1 z_0}$,
since the magnitude of the difference angle will always be the same, regardless of order. 
Given this commutative geometric average, the SLERP operation can also be defined uniquely.

\subsection{Logarithmic Spirals with SLERP}

Finding the shortest paths allows us to use SLERP interpolation in our complex color space. However, SLERP cannot be used for interpolation if one of the numbers is zero as is clear from the SLERP formula given in Equation~\ref{Equ:gi}. The original idea of the SLERP operation was developed to work only in unit quaternion space \cite{shoemake1985animating}, so zero was already excluded. Another problem with using SLERP on the entire complex space is that it produces logarithmic spirals. Let $z_0=r_0 e^{i \theta_0}$ and $z_1=r_1 e^{i \theta_1}$, then
the original SLERP formula given in Equation~\ref{Equ:gi} can be re-written as 
\begin{equation}
z(t) = \left(r_0 e^{i \theta_0}\right) ^{(1-t)} \left(r_1 e^{i \theta_1}\right)^{t} = r_0^{(1-t)}r_{1}^t e^{i ((1-t)\theta_0+t\theta_1))}
\label{Equ:gi2}
\end{equation}
In this equation $r_0^{(1-t)}r_{1}^t$ provides logarithmic scaling and $(1-t)\theta_0+t\theta_1)$ produces rotation. It is clear from this formulation that, even if we avoid zero, small $r$ terms. Also note that since $r$, the distance from the white point corresponds to saturation with $0$ being unsaturated, if we interpolate saturated colors with unsaturated colors the resulting colors will be mostly unsaturated, as can be seen in the upper rows of Figure~\ref{fig:linearinterpolationcomparisonunsaturated}(a) and (b).

\subsection{Archimedean Spirals with CLERP}

\begin{figure}[htb!]
    \begin{subfigure}[t]{0.32\textwidth}
    \includegraphics[width=1.0\textwidth]{6lab.png}
        \includegraphics[width=1.0\textwidth]{6hcl.png}    
        \caption{\it From near white (FFFEFD) to Green (00FF00).}
        \label{fig:Linear/6spiral}
    \end{subfigure}
      \hfill
    \begin{subfigure}[t]{0.32\textwidth}
            \includegraphics[width=1.0\textwidth]{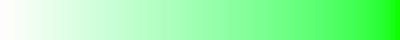}
        \includegraphics[width=1.0\textwidth]{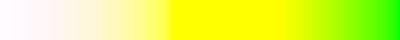}
        \caption{\it From near white (FFFDFE) to Green (00FF00).}
        \label{fig:Linear/7spiral}
    \end{subfigure}
          \hfill
    \begin{subfigure}[t]{0.32\textwidth}
            \includegraphics[width=1.0\textwidth]{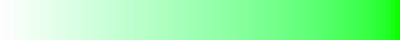}
        \includegraphics[width=1.0\textwidth]{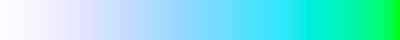}
        \caption{\it From another near white (FEFDFF) to Green (00FF00).}
        \label{fig:Linear/8spiral}
    \end{subfigure}
\caption{Two comparisons of logarithmic and Archimedean spirals interpolating from an unsaturated color to a saturated color. The top is SLERP in LAB and the bottom is circular linear interpolation in LAB.}
\label{fig:linearinterpolationcomparisonunsaturated}
\end{figure}

Now, consider direct linear interpolation in polar forms. Since $r$ corresponds to saturation and $\theta$ corresponds to hue, we can write circular linear interpolation (CLERP) in polar forms as 
\begin{equation}
r = r_0(1-t) + r_1 t \; \; \mbox{and} \; \; 
\theta = \theta_0(1-t) + \theta_1 t\label{Equ:aiHSV}
\end{equation}
Note that CLERP is quite different from SLERP on complex numbers. Since the scaling term changes linearly, unsaturated colors do not dominate the interpolation, as can be seen in the lower rows of Figure~\ref{fig:linearinterpolationcomparisonunsaturated}(a) and (b). Potential hue inconsistencies can also be resolved by first converting control colors into complex space. Then the circular linear interpolation can be done in polar forms to obtain Archimedean spirals. As shown in Figure~\ref{fig:linearinterpolationcomparisonunsaturated}, Archimedean spirals appear significantly better at providing the full gamut of colors. We, therefore, in CLERP we apply SLERP only for angle interpolation. Saturations are still computed by linear interpolation. Note that in this case CLERP is simply linear interpolation in polar forms, but only after computing shortest paths in $\mathbb{C}$. As a result, our comparisons in the rest of the paper can be categorized as barycentric interpolations in three different categories of color spaces, namely, linear in Cartesian  (LERP in RGB), linear in complex  (LERP in CIELAB) linear in polar  (CLERP in CIEHLC).

\section{CLERP: Circular Linear Interpolations in Polar Forms}
\label{sec:LOPF}

The main premise of this approach is to use $\mathbb{C}$ such that properties of different color values such as hue, luminance, chroma, or saturation can be manipulated with operations that can be guaranteed to behave predictably by staying within the convex hull of the colors that serve as operands. In this section, we demonstrate that the complex space $\mathbb{C}$ provides such a formalism that does not require any ad hoc approaches to design operations in any given color space. 

An attractive property of this approach is that there is significant literature on interpolating and approximating curves. We can simply use existing methodologies for color operations without any reformulation and it works without the need for a careful design. The approach can also be extended to other common methodologies such as mean value coordinates. Therefore, in this section, we focus only on linear operations in polar form. 

As we have discussed earlier, linear Operations on CIEHLC partially resemble SLERP operations over complex numbers. We exploit this property to obtain a unique solution by selecting the shortest angular paths. Using shortest paths, we can recompute the angles such that each position can have a unique angle. The pseudo-code is given in Algorithm~\ref{Algol:RecomputingAngles}.

\begin{algorithm}
\caption{Recomputing the angles for an ordered set of complex numbers}
\label{Algol:RecomputingAngles}
\begin{algorithmic}
\REQUIRE $z_k=x_k + i y_k = r_k e^{i \phi_k} $, $\bar{z}_k=x_k - i y_k = r_k e^{-i \phi_k}$, with $k=0,1,\ldots,K$. 
\ENSURE Start with $\phi_0$
\STATE For all $k \in [0,K-1]$:
\STATE \; \; $a+bi \leftarrow z_k \bar{z}_{k+1}$
\STATE  \; \; $ \delta_{k+1} \leftarrow atan2(b,a)$
\STATE \; \; (atan2 function gives the shortest signed angle in $[-\pi, \pi]$)
\STATE  \; \; $ \phi_{k+1} \leftarrow \phi_k + \delta_{k+1} $
\STATE Use the new angles for interpolation
\end{algorithmic}
\end{algorithm}

\begin{figure}[htb!]
     \begin{subfigure}[t]{0.32\textwidth} 
        \includegraphics[width=1.0\textwidth]{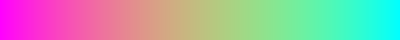}  
        \includegraphics[width=1.0\textwidth]{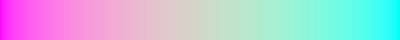}        
         \includegraphics[width=1.0\textwidth]{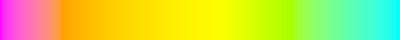}
        \caption{\it FF00FF, FFFF00, 00FFFF.}
        \label{fig:QuadBezier/0}
    \end{subfigure}
      \hfill
    \begin{subfigure}[t]{0.32\textwidth} 
        \includegraphics[width=1.0\textwidth]{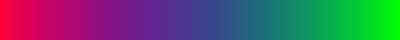}  
        \includegraphics[width=1.0\textwidth]{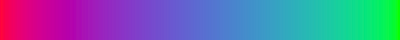}        
         \includegraphics[width=1.0\textwidth]{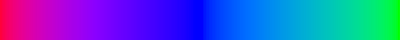}
        \caption{\it FF000F, 0000FF, and 00FF00.}
        \label{fig:QuadBezier/1}
    \end{subfigure}
      \hfill
    \begin{subfigure}[t]{0.32\textwidth}
        \includegraphics[width=1.0\textwidth]{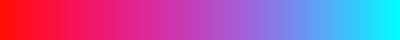}
         \includegraphics[width=1.0\textwidth]{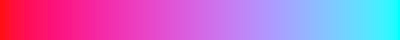}
        \includegraphics[width=1.0\textwidth]{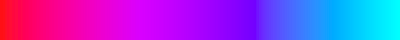}
        \caption{\it FF0F00, FF00FF, and 00FFFF.}
        \label{fig:QuadBezier/2}
    \end{subfigure}
      \hfill
          \begin{subfigure}[t]{0.32\textwidth}
        \includegraphics[width=1.0\textwidth]{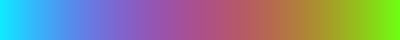}    
        \includegraphics[width=1.0\textwidth]{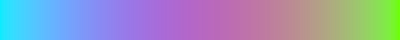}
          \includegraphics[width=1.0\textwidth]{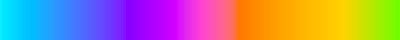}
        \caption{\it 0FEBFF, A00FFF, FF2319, and 6EFF14.}
        \label{fig:CubicBezier/1}
    \end{subfigure}
      \hfill
        \begin{subfigure}[t]{0.32\textwidth}
        \includegraphics[width=1.0\textwidth]{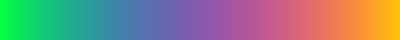}    
        \includegraphics[width=1.0\textwidth]{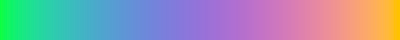}
          \includegraphics[width=1.0\textwidth]{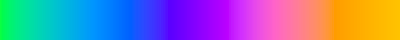}
        \caption{\it 05FF41, 1450FF, FF0AB9, and FFC306.}
        \label{fig:CubicBezier/0}
    \end{subfigure}
      \hfill
        \begin{subfigure}[t]{0.32\textwidth}
        \includegraphics[width=1.0\textwidth]{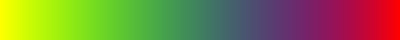}
        \includegraphics[width=1.0\textwidth]{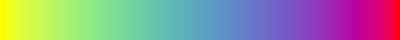}
        \includegraphics[width=1.0\textwidth]{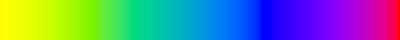}
        \caption{\it FEFF00, 00FF00, 0000FF, and FF0000.}
        \label{fig:CubicBezier/14}
    \end{subfigure}
      \hfill
\caption{Comparison of B\'ezier curves with four different sets of control colors. The top two cases are quadratic B\'ezier, and the bottom two cases are cubic B\'ezier. In each column, the order from top to bottom is RGB, CIELAB, and CIEHLC.
Note that the polar form  CIEHLC comes closer to every control color, while direct interpolations in RGB, and CIELAB miss the middle colors. 
The direct interpolations could be better controlled by using other curves such as beta splines \cite{bartels1995introduction}, rational B\'ezier curves \cite{farin1983algorithms} or NURBS \cite{piegl1996nurbs}.}
\label{fig:BezierExamples}
 \end{figure}

Moving beyond linear interpolation, for open curves $z_0 \neq z_{K}$ and this method will work without a problem. As shown in Figure~\ref{fig:BezierExamples}, the polar forms provide good control to obtain desired results with control points when we use B\'ezier type curves \cite{bartels1995introduction,prautzsch2002bezier} in color spaces. The polar forms do not need such additional tweaks in the shape of the curves.

\section{CAF: Circular Average Filtering}
\label{sec:Filtering}

\begin{figure}
\begin{subfigure}[t]{0.24\textwidth}
    \includegraphics[width=1.0\textwidth]{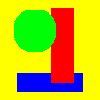}
        \caption{\it Original $100\times 100$ image.}
        \label{fig:filters2/one}
    \end{subfigure}
      \hfill
    \begin{subfigure}[t]{0.24\textwidth}
        \includegraphics[width=1.0\textwidth]{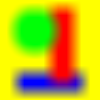}
        \includegraphics[width=1.0\textwidth]{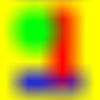}
        \caption{\it RGB.}
        \label{fig:filters2/one15_rgb}
    \end{subfigure}
      \hfill
    \begin{subfigure}[t]{0.24\textwidth}
        \includegraphics[width=1.0\textwidth]{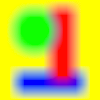}
        \includegraphics[width=1.0\textwidth]{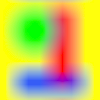}
        \caption{\it   CIELAB.}
        \label{fig:filters2/one15_lab}
    \end{subfigure}
      \hfill
          \begin{subfigure}[t]{0.24\textwidth}
        \includegraphics[width=1.0\textwidth]{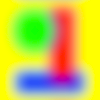}
        \includegraphics[width=1.0\textwidth]{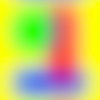}
        \caption{\it  CIEHLC.}
        \label{fig:filters2/one15_hlc}
    \end{subfigure}
\caption{A comparison of linear vs. circular blur filters. The top images are blurred by a $15 \times 15$ box filter and the bottom images by $23 \times 23$. Circular blur in CIELAB corresponds to CIEHLC space and it appears to be best when smoothing.}
\label{fig:onefiltercomparison}
\end{figure}

\begin{figure}
    \begin{subfigure}[t]{0.23\textwidth}
        \includegraphics[width=1.0\textwidth]{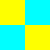}
        \includegraphics[width=1.0\textwidth]{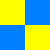}
         \includegraphics[width=1.0\textwidth]{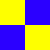}
          \includegraphics[width=1.0\textwidth]{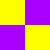}
        \includegraphics[width=1.0\textwidth]{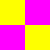}
        \caption{Original 50x50 Image.}
        \label{fig:filters/yb}
    \end{subfigure}
      \hfill
    \begin{subfigure}[t]{0.23\textwidth}
        \includegraphics[width=1.0\textwidth]{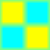}
        \includegraphics[width=1.0\textwidth]{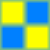}
        \includegraphics[width=1.0\textwidth]{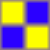}
        \includegraphics[width=1.0\textwidth]{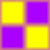}
         \includegraphics[width=1.0\textwidth]{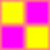}
        \caption{7x7 box filter in RGB.}
        \label{fig:filters/yb_rgb}
    \end{subfigure}
      \hfill
    \begin{subfigure}[t]{0.23\textwidth}
        \includegraphics[width=1.0\textwidth]{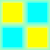}
        \includegraphics[width=1.0\textwidth]{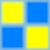}
        \includegraphics[width=1.0\textwidth]{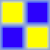}
        \includegraphics[width=1.0\textwidth]{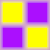}
        \includegraphics[width=1.0\textwidth]{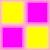}
        \caption{ 7x7 box filter in CIELAB.}
        \label{fig:filters/yb_lab.png}
    \end{subfigure}
         \hfill
        \begin{subfigure}[t]{0.23\textwidth}
        \includegraphics[width=1.0\textwidth]{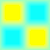}
        \includegraphics[width=1.0\textwidth]{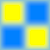}
        \includegraphics[width=1.0\textwidth]{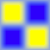}
        \includegraphics[width=1.0\textwidth]{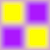}
         \includegraphics[width=1.0\textwidth]{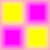}
        \caption{ 7x7 box filter in CIEHLC.}
        \label{fig:filters/yb_hcl.png}
    \end{subfigure}
      \hfill
\caption{Comparison of box filters in variety of colored checkerboard images. In these case, we want you look at artifacts in RGB and CIELAB caused by the sudden change in the boundaries. CIEHLC clearly provides smoother change.}
\label{fig:filtercomparisons}
\end{figure}

For filtering in polar forms, we compute the shortest angular distances from the central pixel and apply the filter to the $\delta$ values to compute an average $\delta$. We then change the color of the central pixel using this average $\delta$ value (See Algorithm~\ref{Algol:RecomputingAngles2}).  These circular average filters in CIEHCL provide very good smoothing as shown in Figures~\ref{fig:filters2/coloredMsm23_hclDetail.png}, ~\ref{fig:filters2/one15_hlc}, \ref{fig:filters/yb_hcl.png}.

\begin{algorithm}
\caption{Recomputing the angles for an ordered set of complex numbers with a given complex number that corresponds to the color of the center pixel.}
\label{Algol:RecomputingAngles2}
\begin{algorithmic}
\REQUIRE $z_0=x_0 + i y_0 = r_0 e^{i \phi_0} $ as representing the color of the center pixel. 
\REQUIRE $z_k=x_k + i y_k = r_k e^{i \phi_k} $, $\bar{z}_k=x_k - i y_k = r_k e^{-i \phi_k}$, with $k=1,\ldots,K$. 
\ENSURE Start with $\phi_0$
\STATE For all $k \in [1,K-1]$:
\STATE \; \; $a+bi \leftarrow z_0 \bar{z}_{k}$
\STATE  \; \; $ \delta_{k} \leftarrow atan2(b,a)$
\STATE \; \; (atan2 function gives the shortest signed angle in $[-\pi, \pi]$)
\STATE  \; \; $ \phi_{k} \leftarrow \phi_k + \delta_{0} $
\STATE Use the new angles for filtering. Apply filter to angles.
\end{algorithmic}
\end{algorithm}

\section{Discussion \& Future Work}

The most important contribution of our paper is the concept of  using the underlying topology of a particular color space to develop an algebraic approach that can work in all color spaces with the same topology. We developed an algebraic approach for operating on color spaces that are homologically similar to CIELAB, which includes CIELUV or CIECAM97s \cite{fairchild2001revision}. On the other hand, we want to point out that there exist color spaces that are homologically different from CIELAB. For instance, the OSA-UCS color space is developed as an approximation of a sphere by using a cuboctahedral-based shape \cite{seim1986towards,huertas2006performance}. Moreover, OSA-UCS allows lightness to be a negative number. This suggests that other algebraic approaches could be based on the space of all 3D rotations, i.e. $SO(3)$, or SU(2), which is isomorphic to unit quaternions.

We also want to point out that RGB is not an appropriate space for algebraization. The projective alpha color model comes close to an algebraic system by representing RGB space with homogeneous coordinates and manipulating them with $4\times4$ matrices \cite{willis2006}. However, the elements are still positive numbers. In order to turn it into vector algebra, we would need to allow all real numbers such that inverses exist. Unlike the chromatic components, in our approach, the value or luminance dimension still lies in the positive real numbers. In our future work, we will examine approaches that allow us to unify all three dimensions in a common algebraic framework.


Similar approaches to the algebraization of color spaces can be useful in a wide variety of applications that do not require a physically based representation of color, such as digital painting, non-photorealistic rendering, and information and scientific visualization. Having an underlying algebraic framework can also help to solve color problems in applications such as digital compositing, where currently ad hoc solutions like negative lights need to be used. Recent work on the non-Riemannian nature of perceptual color space can also provide new avenues \cite{bujack2022non}. 
Similar algebraic approaches can help to provide a rich set of methods for users to design and manipulate colors as in computer-aided geometric design. For instance, subdivision methods can directly be included to produce smooth manifold structures in color spaces. Methodologies such as topological modeling or Morse theory may also find applications to provide additional control to the users in manipulating colors. Multiplication with complex numbers can apply uniform scaling and rotation and it can be used to obtain more controllable illumination effects. Moreover, we can borrow ideas from signal processing to obtain filters with complex coefficients. This could be especially useful to obtain high-pass filters that work directly on the chromatic components of a color.

\bibliographystyle{unsrtnat}
\bibliography{references}  






\end{document}